\documentclass[prl,twocolumn,showpacs,superscriptaddress]{revtex4}

\usepackage{graphicx}
\usepackage{amsmath}
\usepackage{amssymb}
\usepackage{mathrsfs}
\usepackage[british]{babel}

\begin{document}

\title{Ageing effects in single particle trajectory averages}

\author{Johannes Schulz}
\affiliation{Physics Department T30g, Technical University of Munich,
85747 Garching, Germany}
\author{Eli Barkai}
\affiliation{Department of Physics, Bar Ilan University, Ramat-Gan 52900,
Israel}
\author{Ralf Metzler}
\affiliation{Institute for Physics \& Astronomy, University of Potsdam,
14476 Potsdam-Golm, Germany}
\affiliation{Physics Department, Tampere University of Technology,
FI-33101 Tampere, Finland}

\date{\today}

\begin{abstract}
We study time averages of single particle trajectories in scale free
anomalous diffusion processes, in which the measurement starts at some
time $t_a>0$ after initiation of the process at the time origin, $t=0$.
Using ageing renewal theory we show that for such non-stationary processes
a large class of observables are affected by a unique ageing function,
which is independent of boundary conditions or the external forces. We
quantify the weakly non-ergodic nature of this process in terms of the
distribution of time averages and the ergodicity breaking parameter
which both explicitly depend on the ageing time $t_a$. Consequences for the
interpretation of single particle tracking data are discussed.
\end{abstract}


\maketitle

Ergodicity in the Boltzmann sense, the equivalence of sufficiently long time
and ensemble averages of a physical observable, and time translational
invariance are hallmarks of many classical systems. In contrast, ergodicity
violation and ageing effects are found in the dynamics of various complex
systems including glasses \cite{glasses}, blinking quantum dots \cite{qdots},
weakly chaotic maps \cite{maps}, and single particle tracking experiments in
living biological cells \cite{cells}.

As originally pointed out by Bouchaud \cite{glasses}, when the lifetimes $t$ of
states of a physical system are power-law distributed, $\psi(t)\simeq\tau^{
\alpha}/t^{1+\alpha}$ ($0<\alpha<1$), with an infinite mean sojourn time
$\langle t\rangle$, both weak ergodicity breaking and ageing in the following
sense occur: time averages of physical observables remain random even in the
limit of long measurement times and differ from the corresponding ensemble
averages \cite{golan,lubelski,he,deng,stas,neusius,akimoto,heinsalu,irwin,boyer}.
Furthermore ensemble averaged correlation functions of observables, taken at
two time instants $t_2$ and $t_1$, are not solely functions of the time
difference $|t_2-t_1|$. With this breakdown of stationarity, the statistical
properties of such systems are no longer time translation invariant. Consider,
for example, a particle undergoing a random walk in a random environment, such
that the particle hops a finite distance to the left or right with equal
probability, but sojourn times between jumps are distributed like $\psi(t)$.
Such a model, called the continuous time random walk model (CTRW), was
introduced by Scher and Montroll in the context of charge carriers in disordered
systems \cite{montroll}. Its intriguing statistical properties become apparent
when studying time averages, like the time averaged mean squared
displacement (TAMSD), which is commonly used to analyze the diffusive properties
measured in single particle tracking assays. Assume that the process starts at
time $t=0$. From a single trajectory $x(t)$, observed in the time interval
$(t_a,t_a+T)$, the TAMSD is typically determined through the definition
\begin{equation}
\label{intro:tamsd_def}
\overline{\delta^2(\Delta;t_a,T)}=\int_{t_a}^{t_a+T-\Delta}\frac{\left[x(t+
\Delta)-x(t)\right]^2}{T-\Delta}dt,
\end{equation}
with the lag time $\Delta$ and the measurement time $T$.

For Brownian motion, in the limit of large $T$, we obtain the expected behavior,
namely, $\overline{\delta^2}\to2K_1\Delta$. In this case, the process does not
age: the result does not depend on the choice of $t_a$. Moreover Brownian motion
is ergodic: the result for the TAMSD is not random, and the same as found for an
ensemble of Brownian particles, $\langle x^2(\Delta)\rangle=2K_1\Delta$
\cite{REMMM}. Thus the diffusion constant $K_1$ can be determined either from an
ensemble of trajectories as originally done by Perrin \cite{perrin} or, as
conceived by Nordlund \cite{nordlund}, from a single particle trajectory. For
anomalous diffusion, defined in terms of $\langle x^2(\Delta)\rangle=2K_\alpha
\Delta^\alpha/\Gamma(1+\alpha)$ with the generalized diffusion constant $K_{
\alpha}$, the equivalence between time and
ensemble averages as well as the time translation invariance generally breaks
down. One of our central results is that for CTRW processes with long tailed
$\psi(t)$ the aged TAMSD (\ref{intro:tamsd_def}) for $\Delta\ll T$ follows
\begin{equation}
\label{intro:eatamsd}
\left<\overline{\delta^2(\Delta;t_a,T)}\right>=\frac{\Lambda_{\alpha}
(t_a/T)}{\Gamma(1+\alpha)}\frac{g(\Delta)}{T^{1-\alpha}},
\end{equation}
where $\Lambda_\alpha(z)=(1+z)^\alpha-z^\alpha$, and $g$ is a function of
$\Delta$ only. Another important finding is that $\overline{\delta^2}$ remains a
random variable, whose statistical properties explicitly depend on both process
age $t_a$ and measurement time $T$.

Previous work revealed deviations from standard ergodic statistical mechanics
by studying time averages in the interval $(t_a=0,T)$
\cite{lubelski,he,deng,stas,neusius,akimoto,heinsalu}. Namely, in these works
the start of the measurement coincides with the start of the process \cite{REMM}.
However, in experiment, the observed particle may be immersed in the medium long
before we start our observation. In fact in some cases we may not even know when
the process was initiated.

In this Letter, we derive the ageing renewal theory for CTRW processes and
study the dependence of time averages of physical observables on the starting
time $t_a$ of a measurement after system preparation at $t=0$. According to
Eq.~\eqref{intro:eatamsd}, the time intervals $(0,T)$ and $(t_a,t_a+T)$ are not
equivalent: In complete contrast to Brownian motion, the statistical properties
of CTRW trajectories will depend on the observation window, the process exhibits
\emph{ageing}. The remarkable property of Eq.~\eqref{intro:eatamsd} is that
corrections due to ageing enter in terms of a unique prefactor depending on
$t_a$ and the measurement time $T$. We call this $\Lambda_{\alpha}$ the ageing
depression function, which is \emph{independent\/} of $\Delta$. This function
contains all the information on ageing, and is universal since the formula
applies for any external force or boundary condition, and, as shown below,
also holds for  a large class of physical observables. The function $g(\Delta)$
contains the complete $\Delta$-dependence. For instance, we find $g(\Delta)
\simeq\Delta$ for free motion \cite{lubelski,he} and $g(\Delta)\simeq\Delta^{
1-\alpha}$ at long times under confinement \cite{stas,neusius}.

\emph{Ageing renewal theory.} In a CTRW, the position coordinate $x$ of a
random walker is an accumulation of random jump lengths, $x(n)=\sum_{i=0}^n
\delta x_i$. In the simplest, unbiased version of the model, the $\delta x_i$
are independent, identically distributed (IID) random variables with zero mean,
and we assume also finite variance $\sigma^2$. Jumps are separated by random IID
waiting times, drawn from the common distribution $\psi(t)\simeq\tau^\alpha/t^{
1+\alpha}$, with $0<\alpha<1$. This implicitly defines a counting process
$n(t)$, the number of steps up to time $t$. The statistics of the overall
diffusion process $x(t)=x(n(t))$ are to be derived from the properties of both
of its constituents, a
method commonly called subordination \cite{fogedby,weron,friedrich}. For a
large variety of physical applications of CTRW and non-aged renewal theory
see \cite{report,montroll,qd}.
We first discuss the counting process $n(t)$, emphasizing the role of ageing. 

Since waiting times are independent, the counting process $n(t)$ is a renewal
process, which we assume to start at $t=0$. To study the ageing properties
of the system, we consider $n_a(t_a,t)=n(t+t_a)-n(t_a)$, the number of renewals
in the interval $(t_a,t_a+t)$. The corresponding probability density in double
Laplace space, $(t_a,t)\to(s_a,s)$, in the scaling
limit of large times becomes \cite{REM,eli_age}
\begin{equation}
p(n_a;s_a,s)=\delta(n_a)\left(\frac{1}{s_as}-\frac{h(s_a,s)}{s}\right)
+\frac{h(s_a,s)}{s^{1-\alpha}}\tau^\alpha e^{-n_a(s\tau)^\alpha}.
\label{renewal:apdfL}
\end{equation}
Here the probability of the waiting time for the first jump to occur after
start of the measurement at $t_a$ is
\cite{eli_age,koren,godreche}
\begin{equation}
\label{renewal:recL}
h(s_a,s)=\frac{s_a^\alpha-s^\alpha}{s_a^\alpha(s_a-s)}\,\,\,\Leftrightarrow
\,\,\, h(t_a,t)=\frac{t_a^\alpha}{t^\alpha(t_a+t)}.
\end{equation}
In the Brownian limit $\alpha\rightarrow1$, the number of jumps $n$ and real
time $t$ are equivalent, $p(n_a;t_a,t)=\delta(n_a-t/\tau)$.

This formalism allows for a direct calculation of the average of any function
of $n_{a}(t_a,t)$. For instance, the $q$th order moment becomes \cite{supp}
\begin{eqnarray}
\nonumber
\langle n_a^q(s_a,s)\rangle &=& \int_0^\infty n_a^q\, p(n_a;s_a,s) \,dn_a\\
&=&
\Gamma(q+1)\frac{s_a^\alpha-s^\alpha}{s_a^\alpha(s_a-s)}\frac{\tau^\alpha}{s^{
1+\alpha q}}.
\label{renewal:q_1}
\end{eqnarray}
After double Laplace inversion, we find
\begin{eqnarray}
\nonumber
\langle n_a^q(t_a,t)\rangle&=&\Gamma(q+1)/\left[\Gamma(\alpha)\Gamma(1+\alpha
q-\alpha)\right] \hspace{2cm}\\ 
&&\hspace*{-0.8cm}\times\left(\frac{t+t_a}{\tau}\right)^{\alpha q}
B\left(\frac{t}{t+t_a};1+\alpha q-\alpha,\alpha\right),
\label{renewal:q}
\end{eqnarray}
where $B(z;a,b)$ is the incomplete beta function \cite{abramowitz}.
Thus, the number of steps
taken during a time interval of length $t$ is not stationary in distribution:
the moments for the period $[0,t]$ are clearly different from those for $[t_a,
t_a+t]$. Indeed we see from Eqs.~\eqref{renewal:q_1} and \eqref{renewal:q} that
the process gets slower and eventually stalls as $t_a\rightarrow\infty$.
Concurrently the $t$-dependence changes with $t_a$: $\langle n_{a}^q(0,t)\rangle
\simeq t^{\alpha q}$ at $t_a=0$, but $\langle n_{a}^q(t_a,t)\rangle\simeq t_a^{
\alpha-1}t^{1-\alpha+\alpha q}$ for large but finite $t_a/t$.

For the probability density (\ref{renewal:apdfL}) we obtain
\begin{eqnarray}
\nonumber
p(n_a;t_a,t)&\sim&\left[1-m_\alpha(t/t_a)\right]\delta(n_a)
+m_\alpha(t/t_a)\Gamma(2-\alpha)\\
&&\hspace*{-1.0cm} \times\frac{1}{(t/\tau)^{
\alpha}}H^{1,0}_{1,1}\left[\frac{n_a}{(t/\tau)^{\alpha}}\left|\begin{array}{l}(
2-2\alpha,\alpha)\\(0,1)\end{array}\right.\right]
\label{renewal:pdf_lim}
\end{eqnarray}
for $t_a\gg t$, in terms of an $H$-function \cite{mathai,hREM} and the
probability to make at least one step during $[t_a,t_a+t]$,
\begin{equation}
m_{\alpha}(t/t_a)=\frac{B\left([1+t_a/t]^{-1},1-\alpha,\alpha\right)}{
\Gamma(1-\alpha)\Gamma(\alpha)}.
\end{equation}
Again, we emphasize the explicit dependence on $t_a$. When compared to the form
for $t_a=0$ \cite{he,koren}, the most striking difference in
Eq.~(\ref{renewal:pdf_lim}) is the occurrence of the $\delta(n_a)$-term: For a
non-aged process ($t_a=0$), we have $m_{\alpha}=1$, while an
aged process always has a nonzero probability of not performing any steps at
all within the chosen time window, its amplitude approaching 1 algebraically,
$m_{\alpha}\sim(t/t_a)^{1-\alpha}/[\Gamma(\alpha)\Gamma(2-\alpha)]$, as 
$t_a\gg t$.

\emph{Ageing CTRW.} We proceed by adding the position coordinate to our
description. For IID jump distances $\delta x_i$, the jump process $x(n)$
converges in distribution to free Brownian motion in the scaling limit of
many jumps. The anomalous diffusion process $x(t)=x(n(t))$ inherits the
ageing properties of the counting process discussed above. To see this,
consider the $q$th order TA moments
\begin{equation}
\label{ctrw:free:def}
\left<\overline{\delta^q(\Delta;t_a,T)}\right>=\int_{t_a}^{t_a+T-\Delta}\frac{
\langle\left|x(t+\Delta)-x(t)\right|^q\rangle}{T-\Delta}dt,
\end{equation}
which are useful to characterize experimental data \cite{vincent}. For free
Brownian motion $x(n)$ we know that
\begin{equation}
\label{ctrw:free:ea_y}
\langle\left|x(n_2)-x(n_1)\right|^q\rangle=\frac{2\Gamma(q)}{\Gamma(q/2)}
\frac{\sigma^q}{2^{q/2}}|n_2-n_1|^{q/2},
\end{equation}
Since $x(n)$ and $n(t)$ are independent processes, we use conditional averaging
to evaluate the integrand of Eq.~(\ref{ctrw:free:def}). From above result
(\ref{renewal:q}), it follows that
\begin{eqnarray}
\nonumber
\left<\left|x(t+\Delta)-x(t)\right|^q\right>&=&\frac{2\Gamma(q)}{\Gamma(q/2)}
\frac{\sigma^q}{2^{q/2}}\langle n_a^{q/2}(t,\Delta)\rangle\\
\nonumber
&&\hspace*{-3.6cm}
=\Gamma(q+1)/\left[\Gamma(\alpha)\Gamma(1+\alpha q/2-\alpha)\right]\\
&&\hspace*{-3.6cm}
\times\left[K_\alpha\left(t+\Delta\right)^\alpha\right]^{q/2}
B\left(\frac{\Delta}{t+\Delta},1-\alpha+\alpha q/2,\alpha\right),
\label{ctrw:free:ea_x}
\end{eqnarray}
where we identify $K_\alpha=\sigma^2/(2\tau^\alpha)$ \cite{report}. The $q$th
order TA moment (\ref{ctrw:free:def}) in the limit $\Delta\ll T$ then becomes
\begin{equation}
\left<\overline{\delta^q(\Delta;t_a,T)}\right>=\frac{\Lambda_{\alpha}(t_a/T)
\left[K_\alpha\Delta^\alpha\right]^{q/2}\Gamma(q+1)}{\Gamma(\alpha+1)
\Gamma(2-\alpha+\alpha q/2)}\left(\frac{\Delta}{T}\right)^{1-\alpha},
\label{ctrw:free:ta}
\end{equation}
which is of the general shape \eqref{intro:eatamsd} with $g(\Delta)\sim\Delta^{
1-\alpha+\alpha q/2}$. Interestingly, we see the special role of the TAMSD
($q=2$), for which the $\Delta$-scaling is independent of $\alpha$.

\begin{figure}
\includegraphics[width=.5\columnwidth]{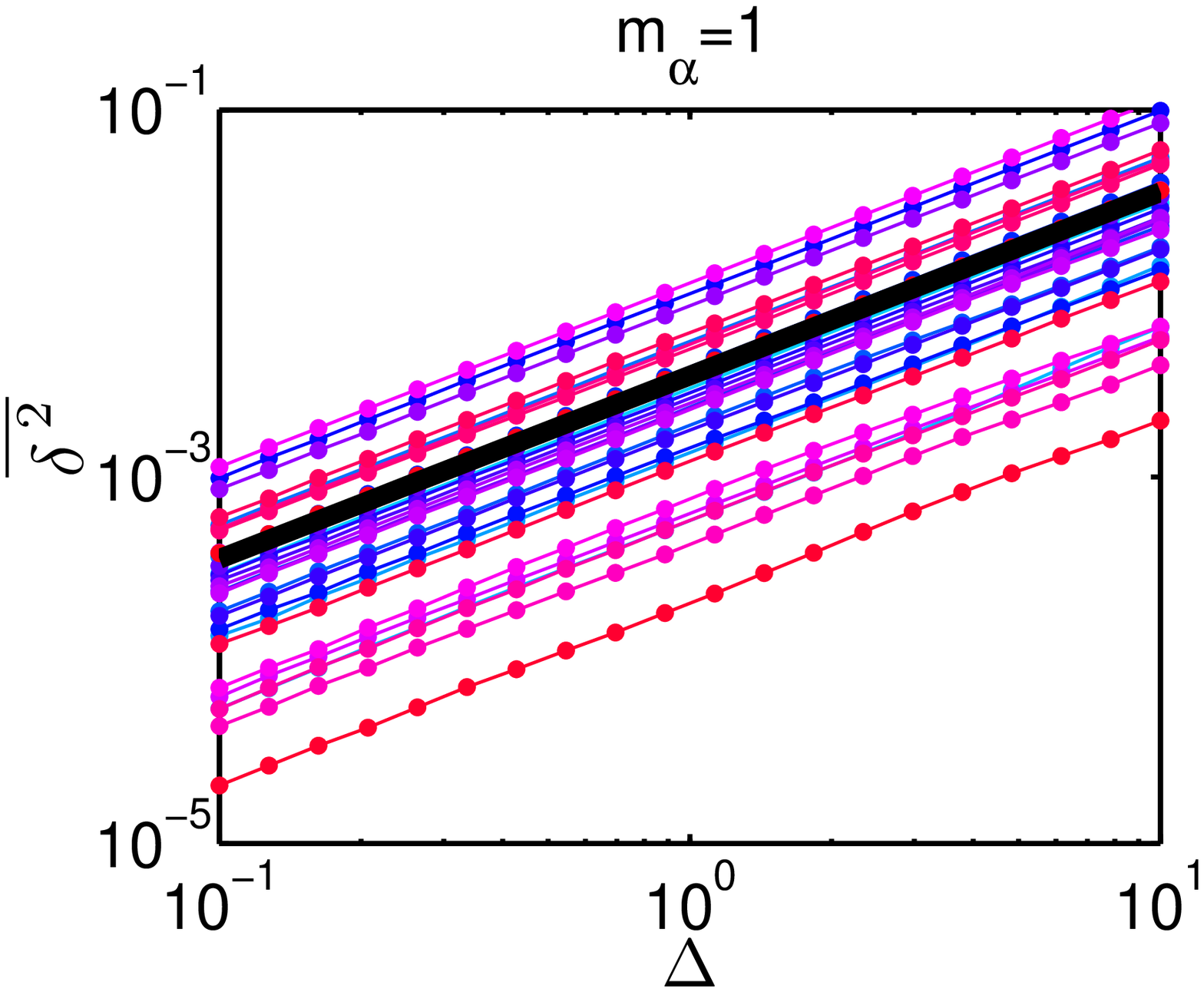}%
\includegraphics[width=.5\columnwidth]{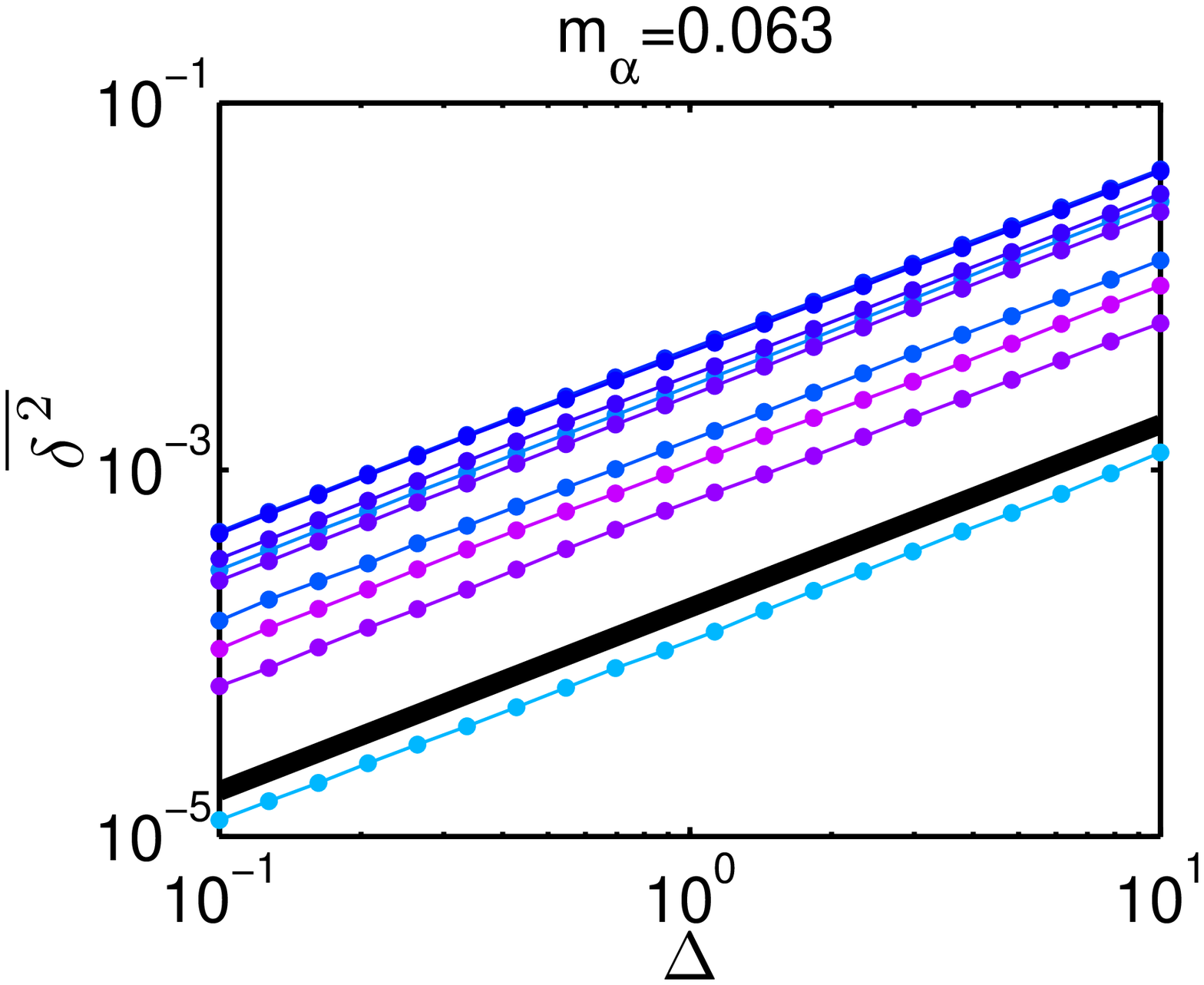}
\caption{Time averaged mean squared displacement $\overline{\delta^2(\Delta;t_a,
T)}$ for individual free CTRW trajectories (full symbols) and the averages
according to Eq.~\eqref{ctrw:free:ta} (bold black lines). Left: Measurement
starts at $t_a=0$, so that $m_\alpha=1$. Right: Aged process, $t_a=10^{11}$
(a.u.), so that a large fraction $1-m_\alpha\approx94\%$ of trajectories is
suppressed in the log-log plot. The parameters are $\alpha=1/2$,
$\tau=1$, $\sigma^2=1$ and $T=10^9$.}
\label{tamsd}
\end{figure}

Fig.~\ref{tamsd} shows simulations results for the TAMSD. If evaluated
during the initial time period, $t_a=0$, the TAMSD for individual trajectories
scatter around the ensemble average $\langle\overline{\delta^2}\rangle$,
Eq.~(\ref{ctrw:free:ta}). In contrast, for the aged process ($t_a\gg
T$) the ensemble average $\langle\overline{
\delta^2}\rangle$ appears much lower than the shown individual trajectories.
This is due to the fact that a significant fraction $1-m_{\alpha}$ of particles
do not move during the entire measurement time. Such trajectories are naturally
not visible in a logarithmic plot, while being included in the
calculation of $\langle\overline{\delta^2}\rangle$.

In a biological cell, the diffusive motion of a tracer particle is spatially
confined, or a charge carrier in a disordered semiconductor experiences a drift
force. To address such systems we now determine the TAMSD in the presence of an
external potential. We start with a Hookean force $-\lambda x$ and address the
more general case below. To that end the continuum approximation of the process
is represented by the Langevin equation
\cite{gardiner}
\begin{equation}
dx/dn=-\lambda x(n)+\xi(n),
\end{equation}
where $\xi(n)$ is white Gaussian noise with $\langle\xi(n_1)\xi(n_2)\rangle=
\sigma^2\delta(n_2-n_1)$. In other words, $x(n)$ is a stationary
Ornstein-Uhlenbeck process. Its increments are Gaussian variables,
characterized by the variance $\langle[x(n_2)-x(n_1)]^2\rangle=
\sigma^2[1-\exp(-\lambda |n_2-n_1|)]/\lambda$. To calculate the TAMSD, we
follow the approach outlined for the free particle, a general approach to
compute this and similar quantities will be provided below. The result reads
\begin{equation}
\left<\overline{\delta^2(\Delta;t_a,T)}\right>=\frac{\Lambda_\alpha(t_a/T)}{
\Gamma(1+\alpha)}\frac{2K_\alpha\Delta}{T^{1-\alpha}}E_{
\alpha,2}\left(-\lambda_{\alpha} \Delta^\alpha\right)
\label{ctrw:bounded}
\end{equation}
in terms of the generalized Mittag-Leffler function \cite{erdelyi}, where
$\lambda_{\alpha}=\lambda/\tau^\alpha$. We deduce the limiting behavior
\begin{equation}
\label{ctrw:bounded:ta}
\left<\overline{\delta^2}\right>\sim\frac{2\Lambda_{\alpha}(t_a/T)K_{
\alpha}}{\Gamma(1+\alpha)T^{1-\alpha}}\left\{\begin{array}{ll}
\Delta, & \Delta\ll\lambda_{\alpha}^{-1/\alpha}\\
\frac{\Delta^{1-\alpha}/\lambda_{\alpha}}{\Gamma(2-\alpha)}, &
\Delta\gg\lambda_{\alpha}^{-1/\alpha}\end{array}\right..
\end{equation}
Eq.~\eqref{ctrw:bounded} is another special case of Eq.~\eqref{intro:eatamsd}.
Interestingly the entire dynamics are multiplied by the unique factor $\Lambda_{
\alpha}$. Let us now test the generality of this feature.

Consider the time average of some observable $F(x_2,x_1)$ along the
trajectory,
\begin{equation}
\label{ctrw:general:def}
\left<\overline{F(\Delta;t_a,T)}\right>=\int_{t_a}^{t_a+T-\Delta}\frac{\langle
F(x(t+\Delta),x(t))\rangle}{T-\Delta}dt.
\end{equation}
$F$ may represent moments ($\overline{F}=\overline{\delta^q}, F(x_2,x_1)=|
x_2-x_1|^q$) or the TA of a correlation function. We only require that the
jump process $x(n)$ and the function $F$ fulfill
\begin{equation}
\langle F(x(n_2),x(n_1))\rangle=f(n_2-n_1).
\label{ctrw:general:ea_y}
\end{equation}
For instance, we have $f(n)=\sigma^2 n$ for the second moment of unbounded
motion
[cf.~\eqref{ctrw:free:ea_y}], or $f(n)=\sigma^2[1-\exp(-\lambda n)]/\lambda$
for the TAMSD in an harmonic potential. Condition \eqref{ctrw:general:ea_y} is
fulfilled whenever $x(n)$ is a stationary process (e.g., equilibrated Brownian
motion). Alternatively, one may consider a process with stationary increments
(e.g., unbounded Brownian motion),
when $F(x_2,x_1)=F(x_2-x_1)$. In these cases we find
\begin{equation}
\label{ctrw:general:cond_av}
\left<\overline{F(\Delta;t_a,T)}\right>=\int_{t_a}^{t_a+T-\Delta}\int_0^\infty
\frac{f(n_a)p(n_a;t,\Delta)}{T-\Delta}dn_adt,
\end{equation}
where $p(n_a;t_a,t)$ is defined in Eq.~\eqref{renewal:apdfL}. We obtain
\begin{equation}
\label{ctrw:general:ta_x}
\left<\overline{F(\Delta;t_a,T)}\right>=C+\frac{\Lambda_{\alpha}(t_a/T)}{
\Gamma(1+\alpha)}\frac{g(\Delta/\tau)}{(T/\tau)^{1-\alpha}},
\end{equation}
at short lag times $\Delta\ll T$, with the constant \(C=f(0)\).
The function $g$ in Laplace space is defined as \cite{REM}
\begin{equation}
g(s)=s^{2\alpha-2}\mathscr{L}\left\{f(n)
-f(0);n\to s^\alpha\right\}.
\end{equation}
Comparing with our specific results \eqref{intro:eatamsd}, \eqref{ctrw:free:ta},
and \eqref{ctrw:bounded:ta}, we identify all relevant terms in TAs:
(i) In the limit $\Delta\to0$, the TA reduces to the constant $C$, which
equals the expectation value of the observable when measured at identical
positions. For example, if we study correlations in an equilibrated
process, $F(x_2,x_1)=x_2x_1$, then $C=\langle x^2\rangle$ is the thermal
value of $x^2$. Conversely, $C$ naturally vanishes if we are interested in
TA moments of displacements, $F(x_2,x_1)=|x_2-x_1|^q$, so it did not appear
previously.
(ii) The lag time dependence
enters through the function $g(\Delta)$. For example, if $f(n)\sim n^q$, then
$C=0$, and in Laplace space $g(s)\sim s^{\alpha-2-\alpha q}$, which implies
$g(\Delta)\sim\Delta^{1-\alpha+\alpha q}$ as in Eq.~\eqref{ctrw:free:ta}. 
(iii) The ageing depression function $\Lambda_{\alpha}$ only depends on the
ratio $t_a/T$ and the parameter $\alpha$, and due to a factor $T^{\alpha-1}$
any TA converges to the constant $C$ as $T\to\infty$. Note that this dependence
on ageing and measurement time $t_a$ and $T$ is universal in the sense that it
is indifferent to the specific choice of observable $F$ or model of the jump
process $x(n)$, but is directly deduced from the nature of the ageing counting
process $n(t)$.
Also note that in the Brownian limit $\alpha=1$, Eq.~(\ref{ctrw:general:ta_x})
reduces to $\langle\overline{F(\Delta;t_a,T)}\rangle=f(\Delta/\tau)$, restoring
the ergodic equivalence of ensemble and time averages, and the stationarity of
the process.

\begin{figure}
\includegraphics[width=\columnwidth]{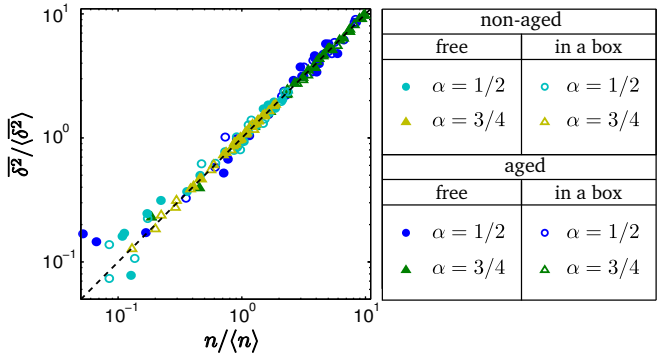}
\caption{Numerical validation of Eq.~\eqref{ctrw:dist:xidef}, for various
$\alpha$, boundary conditions, and $t_a$ (see key). Each point in the
graph represents an individual trajectory. Parameters are $\tau=1$ (a.u.),
$\sigma^2=1$, $\Delta=100$, $T=2\times10^6$. $t_a$ is either $0$
(non-aged), or for specific $\alpha$ chosen such that $m_\alpha=0.054$
(aged).}
\label{fig.nvsta}
\end{figure}

\textit{Distribution of TAMSD.} Due to the scale-free nature of the distribution
$\psi(t)$ of waiting times all TAs of physical observables, e.g. $\overline{
\delta^2}$, remain random quantities, however, with a limiting distribution
$\phi(\xi)$ for the dimensionless ratio $\xi=\overline{\delta^2}/\langle
\overline{\delta^2}\rangle$ \cite{he,stas,jae}. As contributions to time
averages of the form (\ref{intro:tamsd_def}) occur at time instants when the
particle performs a jump, we expect that in the sense of distributions both
$\overline{\delta^2}$ and $n_a$ should be equivalent, $\overline{\delta^2}
\overset{d}{=}cn_a$, for some non-random, positive $c$. In other words,
\begin{equation}
\label{ctrw:dist:xidef}
\xi=\frac{\overline{\delta^2(\Delta;t_a,T)}}{\langle\overline{\delta^2(\Delta;
t_a,T)}\rangle}\overset{d}{=}\frac{n_a(t_a,T)}{\langle n_a(t_a,T)\rangle},
\end{equation}
for $\Delta\ll T$. We may thus deduce the statistics directly from the
underlying counting process. Fig.~\ref{fig.nvsta} provides numerical evidence
for this argument in the case of a free particle and a particle in a box for
several values of $\alpha$.

\begin{figure}
\includegraphics[width=.5\columnwidth]{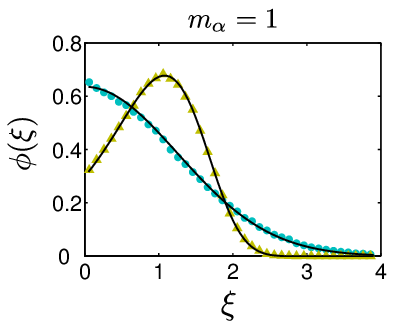}\includegraphics[
width=.5\columnwidth]{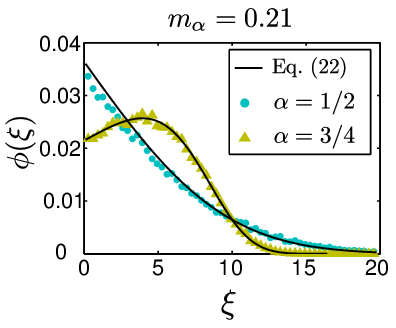}
\caption{Scatter density $\phi(\xi)$ for different $\alpha$ and $m_\alpha$,
see text. Lines: Eq.~(7) from \cite{he} (Left) and Eq.~\eqref{ctrw:dist:dist}
(Right). Symbols: Simulations of free CTRW. Note that the area under the curves
for the aged process in the right panel is not unity, since the fraction $1-
m_{\alpha}$ of immobile events is not shown.
Same parameters as in Fig.~\ref{fig.nvsta}.}
\label{fig.pdf}
\end{figure}

The distribution $\phi(\xi)$ for $t_a=0$ is related to a one-sided L{\'e}vy
stable law
\cite{he}. In the opposite case $t_a\gg T$, combination of
Eqs.~\eqref{ctrw:dist:xidef} and~\eqref{renewal:pdf_lim}, yields
\begin{eqnarray}
\nonumber
\phi(\xi)&\sim&\left[1-m_{\alpha}(T/t_a)\right]\delta(\xi)+m_{\alpha}
(T/t_a)\Gamma(2-
\alpha)\\
&&\hspace*{-1.0cm}\times\frac{\left(T/t_a\right)^{1-\alpha}}{\Gamma(\alpha)}
H^{1,0}_{1,1}\left[\xi\frac{(T/t_a)^{1-\alpha}}{\Gamma(\alpha)}\left|
\begin{array}{ll}(2-2\alpha,\alpha)\\(0,1)\end{array}\right.\right],
\label{ctrw:dist:dist}
\end{eqnarray}
for $\Delta\ll T\ll t_a$. In $\phi(\xi)$, $m_{\alpha}(T/t_a)$ is the weight of
the continuous part. The probability for not moving during the whole measurement
period,
($\xi=0$), approaches unity as $\simeq(T/t_a)^{1-\alpha}$. If conditioned to
measurements with $\xi>0$, we also find the scaling $\xi\sim(t_a/T)^{1-\alpha}$.
In Fig.~\ref{fig.pdf} we demonstrate excellent agreement of
Eq.~(\ref{ctrw:dist:dist}) with numerical simulations.

\begin{figure}
\includegraphics[width=.5\columnwidth]{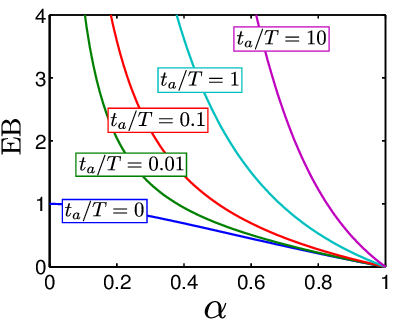}%
\includegraphics[width=.5\columnwidth]{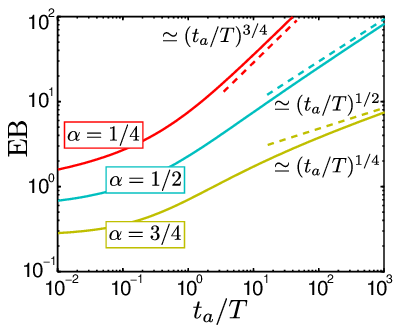}
\caption{Ergodicity breaking parameter (\ref{eb}) as function of $\alpha$
(Left) and $t_a/T$ (Right). Notice that the non-ergodic
fluctuations become larger with increasing $t_a$.}
\label{fig.ebp}
\end{figure}

Deviations from ergodic behavior are quantified by the ergodicity breaking
parameter \cite{he}, for which we obtain
\begin{equation}
\label{eb}
\mathrm{EB}=\frac{\left<\overline{\delta^2}^2\right>}{\left<\overline{\delta^2}
\right>^2}-1=2\alpha\frac{B\left([1+t_a/T]^{-1},1+\alpha,\alpha\right)}{\left[
1-(1+T/t_a)^{-\alpha}\right]^2}-1,
\end{equation}
depending only on the ratio $t_a/T$. At $t_a=0$ EB reduces to the result of
Ref.~\cite{he}, while for $t_a\gg T$, we find $\mathrm{EB}\sim2(t_a/T)^{1-
\alpha}/[\alpha(1+\alpha)]$. In the non-aged case EB is bounded, $0\le\mathrm{
EB}\le1$, depending on the value of $\alpha$ only. In contrast we find that EB
may \emph{diverge\/} in the limit $t_a/T\to\infty$. This implies that the non
ergodic fluctuations are much larger in the aged regime under investigation.
We show the behavior of EB in Fig.~\ref{fig.ebp}.

\emph{Conclusions.}
We investigated the effects of ageing on TAs of physical observables.
Previous
calculations of TAs tacitly neglect the fact that often the preparation
of the system
and start of the measurement do not coincide. While this does not cause
any problems for ergodic systems with rapid memory loss of the initial
conditions, in general this cannot be taken for granted in processes of
anomalous diffusion. Here we showed for the case of CTRW with diverging
characteristic waiting time that TAs of arbitrary physical observables
carry the common factor $\Lambda_{\alpha}$. This ageing depression function
is universal in the sense that it only depends on the process age $t_a$ and
the measurement time $T$. All details such as confinement effects enter through
a single function, $g(\Delta)$. The structure of this result was shown to hold
for a large class of physical observables. We also see that the ageing of the
process has a pronounced statistical effect, splitting the population into two:
the mobile fraction $m_{\alpha}$ and the immobile one whose amplitude $1-
m_{\alpha}$ grows with $t_a/T$.
Knowledge of this effect is significant for the quantitative physical
interpretation of experimental data. Finally, since renewal theory is
applicable to many systems, our results with minor
changes should be relevant more generally, for instance, to the
Aaronson-Darling-Kac theorem in infinite ergodic theory or for counting
the number of renewals in blinking quantum dots.

\acknowledgments

We acknowledge funding from the CompInt graduate school, the Academy of
Finland (FiDiPro scheme), and the Israel Science Foundation.


\begin{thebibliography}{99}
\bibitem{REMMM} In the sense that $\langle x^2(t)\rangle=\int_{-\infty}^{\infty}
x^2P(x,t)dx$, where $P(x,t)$ is the probability density function of the process.
\bibitem{REMM} Note that in some cases the start of the measurement and the
start of the process coincide. For example, in the original work of Scher and
Montroll \cite{montroll}, a light flash excites charge carriers in a disordered
semiconductor. Similarly, in a glassy system we may suddenly quench the
temperature, thus defining physically the start of the process.
\bibitem{REM} We express the Laplace transform $f(s)=\mathscr{L}\{f(t);t\to s\}
=\int_0^{\infty}f(t)\exp(-st)dt$ of a function $f(t)$ by explicit dependence on
the Laplace variable $s$.
\bibitem{hREM} The $H$-function admits series expansions for small and large
arguments which we used to plot the curves in Fig.~\ref{fig.pdf}.
\end{thebibliography}
\end{document}